# Determination of the magnetocaloric effect from thermophysical parameters and their relationships near magnetic phase transition in doped manganites

A.G. Gamzatov[1,*], A.B. Batdalov[1], A.M. Aliev[1], P.D.H. Yen[2], S.V. Gudina[3], V.N. Neverov[3], T.D. Thanh[4], N.T. Dung[4], S.-C. Yu[5], D-H. Kim[2], and M.H. Phan[6,*]

[1]Amirkhanov Institute of Physics, DFRC of RAS, Makhachkala, 367003, Russia

[2]Department of Physics, Chungbuk National University, Cheongju, 28644, South Korea

[3]Miheev Institute of Metal Physics of Ural Branch of RAS, Ekaterinburg, 620108, Russia

[4]Institute of Materials Science, VAST, 18-Hoang Quoc Viet, Hanoi, Vietnam

[5]Ulsan National Institute of Science and Technology, 50, UNIST-gil, Ulju-gun, Ulsan, 44919, South Korea

[6]Department of Physics, University of South Florida, 4202 East Fowler Avenue Tampa, Florida 33620, USA

**Abstract**

We present the results of a comparative analysis of the magnetocaloric effect (MCE) in $Pr_{0.7}Sr_{0.2}Ca_{0.1}MnO_3$, through direct and indirect measurements, using experimentally measured magnetization, specific heat, magnetostriction, resistivity, thermal diffusivity and thermal conductivity parameters. We have demonstrated that the change in each parameter in response to a magnetic field near the ferromagnetic-paramagnetic phase transition temperature of the material correlates with the change in magnetic entropy. These findings allow us to interrelate these parameters and provide an alternative, effective approach for accessing the usefulness of magnetocaloric materials.

*Corresponding authors: gamzatov_adler@mail.ru (A.G), phanm@usf.edu (M.H.P)

**Introduction**

Doped manganites with a general formula $RE_{1-x}AE_x\text{MnO}_3$ ($RE$ = La, Pr, Nd; $AE$ = Ca, Sr, Ba, etc.) represent strongly correlated electron systems in which electron, magnetic and lattice subsystems are interrelated [1-5]. This means that the parameters characterizing the state of one of the subsystems will correlate with those of the other two. Application of external stimuli, such magnetic field, electrical field and pressure, can alter the parameters describing these subsystems (e.g. resistivity, magnetization, specific heat, magnetostriction, etc.), which appear to occur strongly in the vicinity of the magnetic phase transition of the material [1,2,4]. An excellent example of this includes the colossal magnetoresistive (CMR) effect [6], the large magnetocaloric effect (MCE) [7], and the giant magnetostriction (GMS) effect [8], which even co-exist in a doped manganite system.

Recently, an interrelation of the electrical resistivity with the magnetic entropy near a magnetic phase transition has been discussed in doped manganites [9-14]. A simple linear relationship between the magnetic entropy change ($\Delta S_M$) and resistivity with a temperature-independent proportionality coefficient has been established based on the *s-d* exchange models [15] and the double exchange model [16]. The temperature-dependent magnetotransport data were used to determine the magnitude of $\Delta S_M$ with good accuracy, as well as to simulate its temperature dependence around the magnetic phase transition temperature ($T_C$) [9-13]. However, a clear understanding of their association with other parameters, such as magnetostriction and strain, was missing. Our recent study has shown a direct link between the $\Delta S_M$ and the change in volume (the magnetostriction, $\Delta \varepsilon$) of $\text{Pr}_{0.7}\text{Sr}_{0.2}\text{Ca}_{0.1}\text{MnO}_3$ subject to a magnetic field; $\Delta S_M \sim \Delta \varepsilon$ [17].

From the aforementioned studies, it emerges that the change of a spin state in an external magnetic field near $T_C$ not only leads to the appearance of the CMR, MCE, and GMS effects, but also alters electric and thermal transfer processes due to changes in the scattering mechanisms. This appears to be related to the change in spin-phonon interaction affected by a magnetic field near a ferromagnetic-paramagnetic (FM-PM) transition. This thus suggests that changes in resistivity ($\rho$), thermal conductivity ($\kappa$), and thermal diffusivity ($\eta$) in response to the applied magnetic field can be used to qualitatively establish the relationship between the parameters describing the magnetic subsystem and kinetic coefficients. A comparative analysis of these parameters will allow for an evaluation of contributions of different subsystems to the MCE [18]. As $Pr_{0.7}Sr_{0.2}Ca_{0.1}MnO_3$ (PSCMO) undergoes a first-order structural/magnetic transition and possesses the largest $\Delta S_M$ among $Pr_{0.7}Sr_{0.3-x}Ca_xMnO_3$ compositions investigated [19, 20], it is an excellent model system for this purpose of research.

**Results and discussion**

Figure 1 shows the temperature dependence of the specific heat of PSCMO in the temperature range of 80 - 310 K in magnetic fields of 0, 18, and 80 kOe. At room temperature, the specific heat of PSCMO is 106 J/mol K, which is 4.2R. This value agrees well with the high-temperature specific heat values reported previously for analogous manganites [21] and does not contradict with the Dulong – Petit law for a system containing five atoms, including three "light" oxygen atoms. The specific heat obeys the Debye concepts of the specific heat of solids, with exception of a pronounced anomaly at $T_C$ = 196 K, which is associated with a FM-PM phase transition. A sharp peak of the specific heat in a zero field can signal the nature of a first-order transition, but there are no clear signs of thermal hysteresis. In a magnetic field, the peak of the specific heat is suppressed and shifted toward higher temperatures; temperature shift of 17 and 44 K in magnetic fields of 18 and 80 kOe, respectively. This behavior is not characteristic for a second-order FM-PM transition but for a first-order FM-PM transition [22, 23].

The inset in Fig. 1 shows the temperature dependence of the magnetic contribution to the heat capacities $\Delta C_p(T)=C_P-C_{Ph}$ ($C_{Ph}$ – background contribution/lattice contribution) for PSCMO in magnetic fields of 0, 18, and 80 kOe. The magnitude of the specific heat jump in the FM-PM phase transition region is $\Delta C_p \approx 41$ J/mol K, 19 J/mol K, and 5 J/mol K at $H = 0$, 18 and 80 kOe, respectively. The inset in Fig.1 also shows the temperature dependence of the magnetic entropy change associated with the disordering of the magnetic system during the FM-PM phase transition, determined using the formula: $\Delta S'(T) = \int (\Delta C_P/T)dT$. The value of $\Delta S'$ in a zero magnetic field is equal to 2.75 J/mol K, which is much smaller than the ideal values for Ising ($\Delta S'$ = Rln2 = 5.7 J/mol K) and Heisenberg ($\Delta S'$ = Rln4 = 11.52 J/mol K) systems [20]. The difference in the experimentally and theoretically $\Delta S'$ values can arise from an existence of two-phase magnetic state in the material, and therefore, the sample was not completely transformed into a magnetically ordered state, as also discussed in previous studies [21, 23-25]. In a magnetic field, the magnitude of $\Delta S'$ decreases, so the ratio $\Delta S'(H=0)/\Delta S'(H=80$ kOe$) \approx 3$. We should pay attention to the temperature dependence of the entropy change at the magnetic phase transition in a zero magnetic field. Approaching to the transition, starting from $T = 181$ K up to 200 K (a temperature interval: 19 K), the magnitude of $\Delta S'$ increases very rapidly (jump wise) by 8.7 times, and from $T = 200$ K to $T = 253$ K (a temperature interval: 53 K) $\Delta S'$ changes by 1.5 times. In the second temperature regime, a little $\Delta S'(T)$ dependence trend is observed. As a rule, the presence of additional anomalies in the $\Delta S'(T)$ behavior indicates the presence of magnetic or structural inhomogeneities in doped manganites [26]. In this case, however, such an extended transition is likely related to the fact that we are dealing with two combined phase transitions: the structural phase transition (SPT) and the magnetic phase transition (MPT), in which one precedes the other (in this case, SPT + MPT). This is also observed in the $\Delta C_p(T)$ dependence.

Figure 2(a) shows the $\eta$(T) dependence at $H = 0$, 18 and 80 kOe. Let us recall that the thermal diffusivity ($\eta$) is a physical quantity characterizing the rate of establishment of equilibrium temperature distribution in thermal processes [27]. In materials with the prevailing phonon mechanism of thermal conductivity, $\eta$ is related to the phonon mean free path ($l_{ph}$) as $\eta = \frac{1}{3} l_{ph} \vartheta_S$, where $\vartheta_S$ is the speed of sound. Assuming that $\vartheta_S$ is weakly temperature-dependent [28] and taking into account that in doped manganites $\kappa_{ph} \gg \kappa_e$ [29], variation in $\eta$(T) can be determined by $l_{ph}$(T). As can be seen from Fig. 2(a), a sharp decrease in thermal diffusivity accompanied by a deep minimum is observed on the $\eta$(T) near the $T_C$. Several mechanisms may be considered as to explain the emergence of the minimum of $\eta$(T); the removal of Jahn – Teller distortions upon transition to the FM phase [29, 30] and strong spin-phonon scattering near the $T_C$ [28, 31, 32]. A sharp jump in $\eta$(T) usually indicates structural changes. It can thus be assumed that near a magnetostructural phase transition, which is our case, several scattering mechanisms act simultaneously, and the phonon scattering rates by different mechanisms are added, i.e. $\tau_{total}^{-1} = \tau_{ph-JT}^{-1} + \tau_{ph-str}^{-1} + \tau_{ph-m}^{-1}$, (where $\tau_{ph-JT}^{-1}$ - is the phonon scattering rate on Jahn-Teller distortions, $\tau_{ph-str}^{-1}$ - is the phonon scattering rate on structural changes, and $\tau_{ph-m}^{-1}$ is the phonon scattering rate on spin fluctuations). This is the reason for the observed sharp decrease in thermal diffusivity near the $T_C$. The application of an external magnetic field could suppress the phonon scattering related processes, leading to an increase of $\eta$, which is equivalent to an increase of thermal conductivity (Fig. 2 (b)). This can be visualized by constructing the $(\eta_H - \eta_0)/\eta_0 = \Delta\eta/\eta_0$ temperature dependence for different fields (see the inset of Fig. 2a). Firstly, the $\Delta\eta/\eta_0$ value reaches enormous values of 63 and 75 % at $H = 18$ and 80 kOe respectively, which suggests a huge amount of magnetothermal diffusivity in PSCMO. Secondly, on the $\Delta\eta/\eta_0 (T,H)$ dependence at $T \approx 202$ K (according to the M(T) data for this composition, $T_C \sim 202$ K [33]), a fracture is observed, which is apparently due to the change in the dominant scattering mechanisms. The observed negative value of $\Delta\eta/\eta_0$ (~4%) at $T$ = 216 K in $H$ = 18 kOe is probably related to the fact that this magnetic field was not sufficient to completely suppress spin fluctuations.

A similar picture can be drawn out for the magnetic field and temperature dependences of thermal conductivity (Fig. 2b). Upon transition to the FM state at $T$~194 K, thermal conductivity increases by about 40%. The application of 18 or 80 kOe shifts the FM-PM temperature by 10 or 26 K, respectively, leaving the thermal conductivity below the $T_C$. Values of the magneto-thermal conductivity $(\kappa_H - \kappa_0)/\kappa_0 = \Delta\kappa/\kappa_0$ in fields of 18 and 80 kOe are equal to 25 and 32 %, respectively. If we compare values of $\Delta\eta/\eta_0$ and $\Delta\kappa/\kappa_0$ in the corresponding fields, the magnitude of $\Delta\eta/\eta_0$ is more than 2 times greater than that of $\Delta\kappa/\kappa_0$. This is most likely due to the fact that the thermal diffusivity is more sensitive to the magnetostructural change than the thermal conductivity.

Figure 3 shows the $\rho(T)$ dependence for PSCMO in the temperature range of 70 - 380 K in magnetic fields of 0 and 80 kOe. The dependence of the resistivity on temperature and magnetic field for PSCMO shows a typical characteristic of doped manganites. The left of Fig. 3 shows the $\Delta\rho/\rho(T)$ dependence for $H$ = 80 kOe. The $\Delta\rho/\rho$ change in 80 kOe is equal to 91%. As we recall, a direct relationship between the magnetic entropy/ MCE and the resistivity/ magnetoresistance (MR) has been reported in doped manganites near their Curie temperatures [11-14]. In the present study, the MCE magnitude can also be estimated from the MR data, following the method [13, 14], and the results of which are presented and discussed below.

The $\Delta S_M$ of PSCMO has been estimated using both direct and indirect measurements. From Maxwell's relation $\mu_0(\partial M/\partial T)_H = (\partial S/\partial H)_T$ [33], the $\Delta S_M$ can be determined as

$$\Delta S_M = \mu_0 \int_0^H \left(\frac{\partial M}{\partial T}\right)_H dH \qquad (1)$$

With the adiabatic temperature change, $\Delta T_{ad}$, obtained from direct measurements and the experimentally measured specific heat $C_P(T, H)$, the $\Delta S_M$ can be determined as

$$\Delta S_M = -\Delta T_{ad} \frac{C_P(T,H)}{T} \qquad (2)$$

From the temperature-dependent specific heat at $H = 0$ and $H \neq 0$, the $\Delta S_M$ can also be determined as

$$\Delta S_M = \int_{T_1}^{T_2} ((C_P(T, H_0) - C_P(T, H_1))/T)_{P,H} dT \qquad (3)$$

In addition to the aforementioned formulas for assessing $\Delta S_M$, there are other indirect methods. As was stated in the introduction, $\Delta S_M$ can also be estimated from the resistivity or magnetoresistive data using the following expressions [12, 13]:

$$\Delta S_M^\rho = -A(MR_{total} - TMR), \qquad (4a)$$

$$\Delta S_M^\rho = -\alpha \int_0^H \left[\frac{\partial ln\rho}{\partial T}\right]_H dH \qquad (4b)$$

where parameters $A$ and $\alpha$ determine the magnetic property of a substance ($A \approx$ 14-15 J/kg K [13, 14], $\alpha \approx$ 21 emu/g [12], and $MR_{total}$ and $TMR$ are the total and tunneling magnetoresistance, respectively.

It is generally known that the occurrence of a first-order magnetic phase transition is often accompanied by a considerable change in the lattice volume, which in turn leads to the change in magnetization of the sample, thus contributing to the total entropy change [18]. It has been shown that for PSCMO, the $\Delta S_M$ can be estimated from the magnetostriction data (in the case magnetostriction near the $T_C$ shows an isotropic behavior, i.e., the volume magnetostriction $\Delta\varepsilon = \lambda_{||} + 2\lambda_\perp \approx 3\lambda_{||}$) using a simple ratio [17]:

$$\Delta S_M^\varepsilon = -\gamma \Delta\varepsilon(T, H), \qquad (5)$$

where the parameter $\gamma$ is independent of the magnetic field and is $2 \cdot 10^4$ J/kg K for PSCMO. $\Delta\varepsilon$ is the magnetostriction.

In one hand, the MCE is due to a change in the degree of ordering of the magnetic subsystem when the applied magnetic field is changed. On the other hand, if the material dissipates the heat carriers on the spin system, then the change in the degree of ordering of the magnetic subsystem will lead to changes in the thermal diffusivity (see Fig. 2) and the thermal conductivity. By considering the relation $\kappa = dC_P\eta$ (where $\kappa$ is the thermal conductivity, $d$ is the density), i.e. $C_P \sim 1/\eta$, which relates the thermal diffusivity to the specific heat, we can qualitatively estimate the $\Delta S_M$ of the system from the $\eta(T, H)$ data using the following expression [20]:

$$\Delta S_\eta = -B \int_{T_1}^{T_2} \frac{((1/\eta(T,H_0))-(1/\eta(T,H_1)))}{T} dT, \quad (6)$$

where $B$ is a coefficient independent of temperature and magnetic field.

Paying attention to the $\Delta\kappa/\kappa(T)$ dependence at $H = 18$ and 80 kOe (see inset in Fig. 2 (b)), one can see that temperature dependences of $\Delta\kappa/\kappa$ and $\Delta S_M$ [17] are quite identical in nature. This means that in doped manganites, the change in thermal conductivity in response to a magnetic field near $T_C$ can also be related to the change in magnetic entropy: $\Delta\kappa/\kappa(T) \sim \Delta S_M$. Accordingly, another indirect method can be used for estimating the MCE in manganites using:

$$\Delta S_M^\kappa = -q \frac{\Delta\kappa}{\kappa_0}(T, H), \quad (7)$$

where $q \approx 25.71$ J/kg K is the proportionality coefficient independent of temperature and magnetic field. The values of $\Delta S_M$ calculated using Eqns. (1)-(7) for a magnetic field change of 80 kOe are presented in Fig. 4. The inset in Fig. 4 shows an estimate of $\Delta S_M$ using Eqns. (1)-(3) and (6) for a magnetic field change of 18 kOe.

We have used the following Eqn. (4a) to estimate $\Delta S_M$ using the resistivity data. As is known, in polycrystalline manganites, in addition to the classical MR effect $(\Delta\rho/\rho)_{sd}$, another TMR effect, based on the intergranular tunneling of the charge carriers through the grain boundaries, may become significant, especially at low temperatures, thus contributing to the total magnetoresistance $(\Delta\rho/\rho)_{total}$. The use of Eqns. (4a)-(4b) for estimating the MCE from the resistivity and magnetoresistance data without taking these factors into account could be erroneous. Therefore, to separate the various contributions to the total MR, we will adhere to the ideas proposed in [13, 14]. As for the TMR description, we will use the formula proposed in works [34, 35]:

$$TMR(H,T) = -P^2 L^2 \left[ \frac{\mu_g H \left(\frac{b}{T}\right)^{3/2}}{kT} \right], \quad (8)$$

where $P$ is polarization, $L(x)$ is the Langevin function ($L(x) = \coth(x) - 1/x(x)$), $\mu_g$ and $b$ is the fitting parameters associated with the magnetic moment and the size of the granules, respectively. In Fig. 3, the dotted line denominates Eqn. (8) for TMR at $H = 80$ kOe (with the parameters for approximating $P = 0.7033$, $b = 0.604$ and $\mu = 2.18 \cdot 10^{-21}$).

$\Delta S_M$ values calculated from the $M(T,H)$ data for this sample were taken from [19] (for a field change of 80 kOe). The $\Delta S_M(T)$ values were obtained by building $\Delta S_M(H)$ curves at different temperatures and extrapolating them to 80 kOe. As we can see, the $\Delta S_M(T)$ values, obtained from Eqns. (1) and (2) (used the $M(T,H)$ and $\Delta T_{ad}(T)$ data, respectively) for a field change of 80 kOe are almost identical. These values $of$ $\Delta S_M$ are also close to those estimated from the magnetostriction data [Eqn. (5)] and thermal conductivity data [Eqn. (7)]. In general, the values of $\Delta S_M$ and their temperature dependence estimated from indirect and direct measurements are in good agreement in the ferromagnetic region. It should be noted that the $\Delta S_M$ values obtained according to Eqns. (1)-(7) for a filed change of 18 kOe provide a quantitative picture (see the inset of Fig. 4). The difference in the MCE maximum temperatures obtained from these two different estimation methods (see Table 1 and Fig. 4) can be related to the existence of magnetic inhomogeneities within the sample with a first-order phase transition; the $T_C$ distribution of which may be of Gaussian or other nature [36].

The noticeable differences in $\Delta S_M(T)$ estimated from the thermal diffusivity data are observed at $T > T_C$ in high magnetic fields (80 kOe). There could be several reasons for this discrepancy. Firstly, it can be connected with a strong shift of the critical temperature in the magnetic field, which is typical for a magneto-structural phase transition (by about 30 ~ 40 K in the field of 80 kOe). The differences in $\Delta S_M(T)$ at $T > T_C$, obtained from $C_P(T, H)$ and $\rho(T, H)$ data, are due to this effect as well. Secondly, as we mentioned above, near the $T_C$ of the material, the scattering of heat carriers could occur mainly on spin fluctuations, on local distortions of the crystal lattice (Jahn-Teller distortions), on structural defects due to the strong change in the volume of the lattice upon the structural phase transition. The total effect of these scattering contributions results in a deep minimum in the $\eta(T)$ dependence. Upon estimating $\Delta S_M(T)$ by the $\eta(T)$ data, the changes in contributions to the magnetic field are especially important. In low magnetic fields (up to 20 kOe), the structural contribution prevails [10], therefore we observe a qualitative agreement of $\Delta S_\eta$ in a field of 18 kOe (estimated by Eqn. (6), see the inset of Fig.4) with the standard caloric estimates (from Eqns. (1)-(3)). With large magnetic fields applied, other mechanisms of heat scattering already dominate, changes in the magnetic field of which do not greatly affect the magnitude of $\Delta S_M$. As a result of the changes in the magnetic field-influenced scattering mechanisms, the $\Delta S_\eta(T)$ dependence at $T > T_C$ (at $H$ = 80 kOe) does not fit the standard behavior of $\Delta S_M(T)$, but the $\Delta S_\eta^{max} \approx 8$ J/kg K is consistent with the data obtained from caloric measurements (see Table 1). In low magnetic fields (< 20 kOe), other indirect estimates of $\Delta S_M(T)$ from Eqns. (4) - (7) also provide a good agreement with those obtained from the standard caloric measurements (see Table 1).

**Conclusion**

We have estimated the magnetic entropy change ($\Delta S_M$) of PSCMO using means of the isothermal magnetization $M$(T,H), the specific heat $C_P$(T,H) and $\Delta T_{ad}$(H,T), the magnetostriction $\Delta\varepsilon$(H,T), the resistivity $\rho$(H,T), the thermal diffusivity $\eta$(H,T), and the thermal conductivity $\kappa$(H,T). The results indicate that the mechanisms leading to the magnetic field-dependent thermophysical parameters near the $T_C$ are directly connected with the spin-phonon interaction. The assessable MCE, using Eqns. (1)-(7), asserts that regularities in doped manganites near the second-order phase transition in a certain range of magnetic fields are universal and can be formulated as follows: *an alteration of $\Delta S_M$ near $T_C$ is directly proportional to the change in $F(T,H)$ in response to a magnetic field, i.e. $\Delta S_M \sim \mp \Delta F(T,H) = F(T,H=0) - F(T,H \neq 0)$*. $F(T,H)$ can be represented by the experimentally measured thermodynamic, kinetic and thermophysical parameters. The sign "+" or "-" depends on the influence of the magnetic field on the measured coefficients.

**Acknowledgements**

The work was funded by RFBR and VAST under Grant number 20-58-54006 and by the Russian Science Foundation under Grant number 18-12-00415. The magnetoresistance measurements were carried out at the Collaborative Access Center «Testing Center of Nanotechnology and Advanced Materials» of M.N. Mikheev Institute of Metal Physics of the Ural Branch of the Russian Academy of Sciences. The work was carried out as part of the state task of the Ministry of Science of the Russian Federation (# AAAA-A17-117021310366-5).

**References**

1. E. Dagotto, Complexity in strongly correlated electronic systems, *Science*, **309**, 257 (2005).

2. E. Dagotto, Open questions in CMR manganites, relevance of clustered states and analogies with other compounds including the cuprates, *New Journal of Physics* **7**, 67, (2005).

3. Y. Tokura and N. Nagaosa, Orbital Physics in Transition-Metal Oxides, *Science* **288**, 462 (2000).

4. M. Kim, H. Barath, X. Chen, Y.I. Joe, E. Fradkin, P. Abbamonte, S.L. Cooper, Magnetic-field- and pressure-induced quantum phases in complex materials, *Advanced Materials* **22**, 1148 (2010).

5. V. Markovich, A. Wisniewski, H. Szymczak, Magnetic Properties of Perovskite Manganites and Their Modifications, *Handbook of Magnetic Materials*, Edited by K.H.J. Buschow, 22, 1-201 (2014)

6. Y. Tokura, Critical features of colossal magnetoresistive manganites, *Reports on Progress in Physics* **69**, 797 (2006).

7. M. H. Phan, and S. C. Yu, Review of the magnetocaloric effect in manganite materials, *Journal of Magnetism and Magnetic Materials* **308**, 325 (2007).

8. R. Mahendiran, M.R. Ibarra, C. Marquina, B. Garcia-Landa, and L. Morellon, Giant anisotropic magnetostriction in $Pr_{0.5}Sr_{0.5}MnO_3$, *Appl. Phys. Lett*. **82**, 242 (2003).

9. R. Rawat, I. Das, The similar dependence of the magnetocaloric effect and magnetoresistance in TmCu and TmAg compounds and its implications, *J. Phys.: Cond. Matter* **13**, L379 (2001).

10. N. Sakamoto, T. Kyômen, S. Tsubouchi, and M. Itoh, Proportional relation between magnetoresistance and entropy suppression due to magnetic field in metallic ferromagnets, *Physical Review B* **69**, 092401 (2004).

11. J.C.P. Campoy, E.J.R. Plaza, A.A. Coelho, S. Gama, Magnetoresistivity as a probe to the field-induced change of magnetic entropy in RAl2 compounds (R=Pr, Nd, Tb, Dy, Ho, Er) *Physical Review B* **74**, 134 410 (2006).

12. C.M. Xiong, J.R. Sun, Y.F. Chen, B.G. Shen, J. Du, Y.X. Li, Relation between magnetic entropy and resistivity in $La_{0.67}Ca_{0.33}MnO_3$, *IEEE Transactions on Magnetics* **41**, 122-124 (2005).

13. A.G. Gamzatov, A.B. Batdalov, The relation between magnetoresistance and magnetocaloric effect in $La_{0.85}Ag_{0.15}MnO_3$ manganite, *Physica B* **406**, 1902-1905 (2011).

14. A. G. Gamzatov and I. K. Kamilov, Tunneling magnetoresistance and indirect measurement of the magnetocaloric effect in lanthanum deficient manganite $La_{0.8}Ag_{0.1}MnO_3$, *Journal of Applied Physics* **114**, 093902 (2013).

15. T. Kasuya, Electrical resistance of ferromagnetic metals, *Prog. Theor. Phys.* **16**, 58 (1956).

16. K. Kubo, N. Ohata, A quantum theory of double exchange, *Journal of the Physical Society of Japan* **33**, 21 (1972).

17. A. G. Gamzatov, A. M. Aliev, P. D. H. Yen, L. Khanov, K. X. Hau, T. D. Thanh, N. T. Dung, and S.-C. Yu, Correlation of the magnetocaloric effect and magnetostriction near the first-order phase transition in $Pr_{0.7}Sr_{0.2}Ca_{0.1}MnO_3$ manganite, *Journal of Applied Physics* **124**, 183902 (2018).

18. A. M. Aliev, A. B. Batdalov, and L. N. Khanov, Magnetic and lattice contributions to the magnetocaloric effect in $Sm_{1-x}Sr_xMnO_3$ manganites, *Applied Physics Letters* 112, 142407 (2018).

19. Y. Pham, T. D. Thanh, T. V. Manh, N. T. Dung, W. H. Shon, J. S. Rhyee, D.-H. Kim and S.-C. Yu, Magnetocaloric effect and the change from first- to second-order magnetic phase transition in $Pr_{0.7}Ca_xSr_{0.3-x}MnO_3$ polycrystalline compounds, *AIP Advances* **8**, 101417 (2018).

20. A.B. Batdalov, A.G. Gamzatov, A.M. Aliev, N. Abdulkadirov, P.D.H. Yen, T.D. Thanh, N.T. Dung, S.-C. Yu, Magnetocaloric properties in the $Pr_{0.7}Sr_{0.3-x}Ca_xMnO_3$: Direct and indirect estimations from thermal diffusivity data, *Journal of Alloys and Compounds* **782**, 729-734 (2019).

21. M. R. Lees, O. A. Petrenko, G. Balakrishnan, and D. McK. Paul, Specific heat of $Pr_{0.6}(Ca_{1-x}Sr_x)_{0.4}MnO_3$ (0<x<1), *Physical Review B* **59**, 1298 (1999).

22. P. Lin, S. H. Chun, M.B. Salamon, Y. Tomioka, Y. Tokura, Magnetic specific heat in lanthanum manganite single crystals, *Journal of Applied Physics* **87**, 5825 (2000).

23. B. F. Woodfield, M. L. Wilson, and J. M. Byers, Low-Temperature Specific Heat of $La_{1-x}Sr_xMnO_{3+\delta}$, *Physical Review Letters* **78**, 3201-3204 (1997).

24. Y. Moritomo, A. Machida, E. Nishibori, M. Takata, and M. Sakata, Enhanced specific heat jump in electron-doped $CaMnO_3$: Spin ordering driven by charge separation, *Physical Review B* **64**, 214409 (2001).

25. A. G. Gamzatov, A. B. Batdalov, A. M. Aliev, Z. Khurshilova, M. Ellouze, F. Ben Jemma, Specific heat, thermal diffusion, thermal conductivity and magnetocaloric effect in Pr0.6Sr0.4Mn1-xFexO3 manganites, *Journal of Magnetism and Magnetic Materials* **443**, 352-357 (2017).

26. V. Franco, A. Conde, Scaling laws for the magnetocaloric effect in second order phase transitions: From physics to applications for the characterization of materials, *International Journal of Refrigeration* **33**, 465-473 (2010).

27. R. Berman, Thermal Conduction in Solids (Oxford Studies in Physics), Oxford University Press, 206 P. (1976).

28. M. Ikebe, H. Fujishiro and Y. Konno, Anomalous Phonon-Spin Scattering in $La_{1-x}Sr_xMnO_3$, *Journal of the Physical Society of Japan* **67**, 1083-1085 (1998).

29. J. L. Cohn, J. J. Neumeier, C. P. Popoviciu, K. J. McClellan, and Th. Leventouri, Local lattice distortions and thermal transport in perovskite manganites, *Physical Review B* **56**, R8495 (1997).

30. D. W. Visser, A. P. Ramirez, and M. A. Subramanian, Thermal conductivity of manganite perovskites: colossal magnetoresistance as a lattice-dynamics transition, *Phys. Rev. Lett.* **78**, 3947 (1997).

31. M. Tachibana and E. Takayama-Muromachi. Thermal conductivity of colossal magnetoresistive manganites $(La_{1-x}Nd_x)_{0.7}Pb_{0.3}MnO_3$, *Applied Physics Letters* **92**, 242507 (2008).

32. A.B. Batdalov, A.G. Gamzatov, A.M. Aliev, L.N. Khanov, A.A. Mukhuchev, I.K. Kamilov, Mechanisms of heat carriers scattering in $La_{1-x}Sr_xMnO_3$ single crystals near the phase transition temperature, *Journal of Alloys and Compounds* **705**, 740-744 (2017).

33. A. M. Tishin and Y. I. Spichkin, The Magnetocaloric Effect and its Applications Institute of Physics, New York, (2003).

34. E. Z. Meilikhov, Tunneling magnetoresistance and Hall effect of granular ferromagnetic metals, *JETP Lett.* **69**, 623 (1999).

35. S. Honda, T. Okada, M. Nawate, and M. Tokumoto, Tunneling giant magnetoresistance in heterogeneous Fe−SiO$_2$ granular films, *Physical Review B* **56**, 14566 (1997).

36. N. G. Bebenin, R. I. Zainullina, V. V. Ustinov, Y. M. Mukovskii, Effect of inhomogeneity on magnetic, magnetocaloric, and magnetotransport properties of $La_{0.6}Pr_{0.1}Ca_{0.3}MnO_3$ single crystal, *Journal of Magnetism and Magnetic Materials*, **324**(6), 1112-1116 (2012).

**Figure captions**

**Fig. 1.** Temperature dependence of the specific heat for PSCMO at $H = 0$, 18 and 80 kOe. Inset shows the temperature dependence of the magnetic contribution of $\Delta C_P$ (solid points) and the change in magnetic entropy $\Delta S'$ (solid lines) at $H = 0$, 18 and 80 kOe.

**Fig. 2.** a) Temperature dependence of the thermal diffusivity for PSCMO at $H = 0$, 18 and 80 kOe. The inset shows the temperature dependence $\Delta \eta / \eta_0$ at $H = 18$ and 80 kOe. b) Temperature dependence of the thermal conductivity for PSCMO at $H = 0$, 18 and 80 kOe. The inset shows the temperature dependence of $\Delta \kappa / \kappa_0$ at $H = 18$ and 80 kOe.

**Fig. 3.** Temperature dependence of the resistivity in magnetic fields $H = 0$ and 80 kOe; The temperature dependence of magnetoresistance in $H = 80$ kOe is also plotted.

**Fig. 4.** Temperature dependences of $\Delta S_M$, obtained using different methods for a field change of 80 kOe. The inset shows the $\Delta S_M(T)$ data for $H = 18$ kOe.

**Fig. 1**

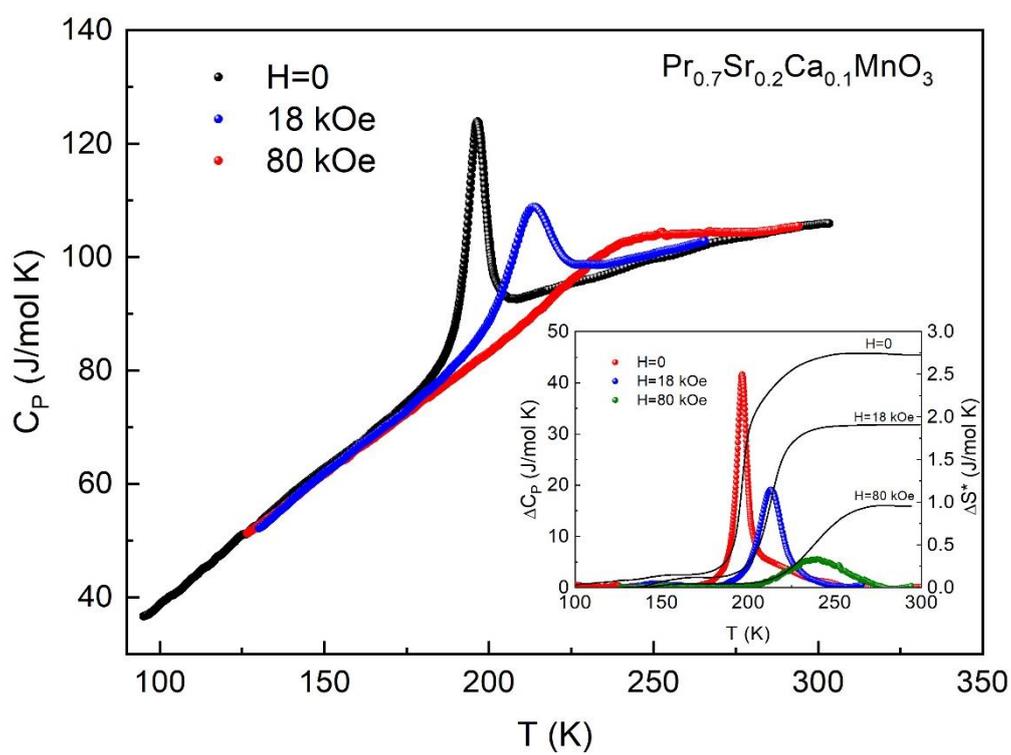

**Fig. 2**

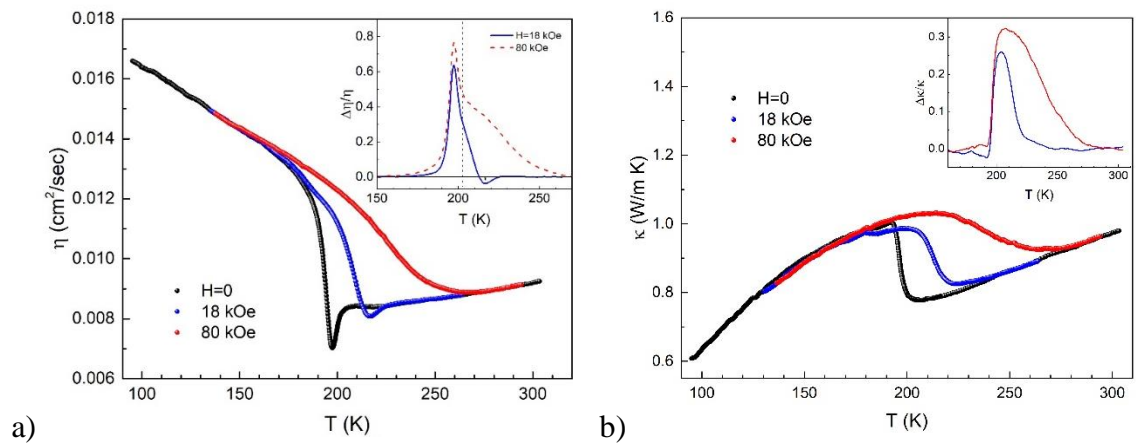

**Fig. 3**

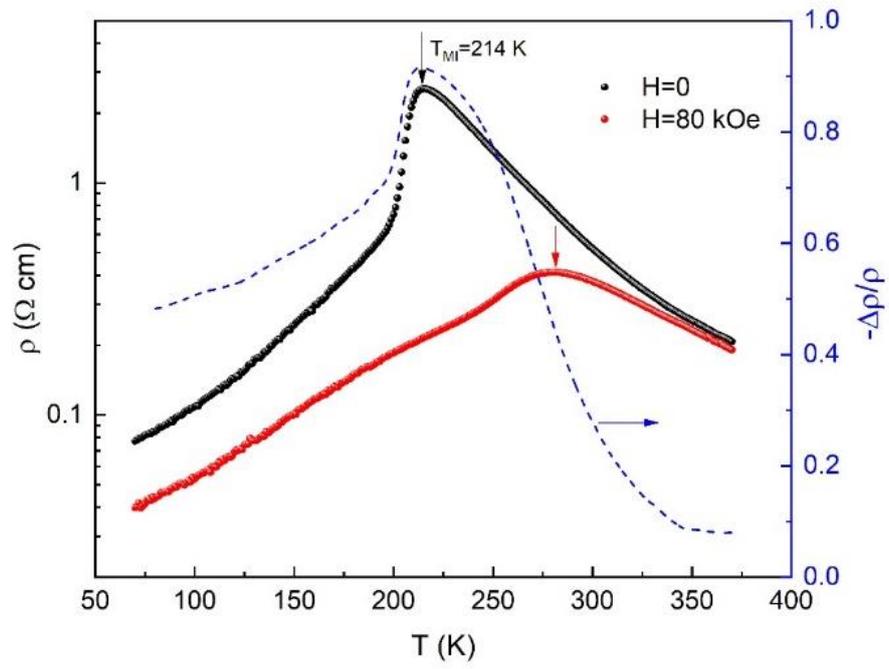

**Fig. 4**

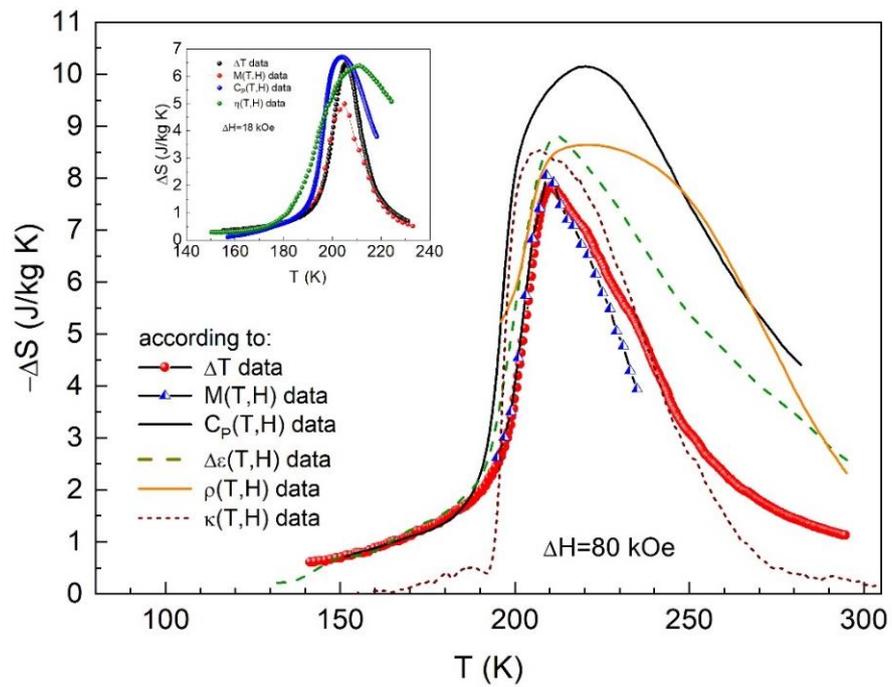

**Table 1.** Some characteristics of the sample

| According to data | $M(T,H)$ | $C_P(T,H)$ | $\Delta T(H,T)$ | $\Delta\varepsilon(H,T)$ | $\rho(H,T)$ | $\eta(H,T)$ | $\kappa(H,T)$ |
|---|---|---|---|---|---|---|---|
| $\Delta S_M$, J/kg K $H = 80$ kOe | 8.05 | 10.15 | 7.88 | 8.79 | 8.65 | 8.23 | 8.51 |
| $T_{max}$ ($H$=80 kOe) | 208 | 219 | 210 | 212 | 221 | - | 207 |
| RCP, J/kg K | 282 | 813 | 328 | 554 | 709 | - | 348 |
| $\Delta S_M$, J/kg K $H$=18 kOe | 4.99 | 6.75 | 6.52 | - | - | 6.41 | - |
| $T_{max}$ ($H$=18 kOe) | 205 | 203 | 205 | - | - | 210 | - |